# Automated Road Safety: Enhancing Sign and Surface Damage Detection with AI

**Davide Merolla[1], Vittorio Latorre[1], Antonio Salis[2], Gianluca Boanelli[3]**

[1] Università degli Studi del Molise, Contrada Fonte Lappone – I-86090 Pesche (IS), Italy
{davide.merolla, vittorio.latorre}@unimol.it

[2]Tiscali Italia S.p.A. Loc. Sa Illetta, SS195 km 2,300 – I-09122 Cagliari, Italy
{antonio.salis}@tiscali.com

[3]Tiscali Italia S.p.A. V.le Città d'Europa, 681, I-00144 Roma, Italy
{gianluca.boanelli}@tiscali.com

**Abstract.** Public transportation plays a crucial role in our lives, and the road network is a vital component in the implementation of smart cities. Recent advancements in AI have enabled the development of advanced monitoring systems capable of detecting anomalies in road surfaces and road signs, which, if unaddressed, can lead to serious road accidents. This paper presents an innovative approach to enhance road safety through the detection and classification of traffic signs and road surface damage using advanced deep learning techniques. This integrated approach supports proactive maintenance strategies, improving road safety and resource allocation for the Molise region and the city of Campobasso. The resulting system, developed as part of the Casa delle Tecnologie Emergenti (House of Emergent Technologies) Molise (Molise CTE) research project funded by the Italian Minister of Economic Growth (MIMIT), leverages cutting-edge technologies such as Cloud Computing and High Performance Computing with GPU utilization. It serves as a valuable tool for municipalities, enabling quick detection of anomalies and the prompt organization of maintenance operations.

## 1. Introduction

According to latest World Health Organization (WHO) [1] public road networks are the lifeblood of modern societies, playing a crucial role for goods and people transportation, which is fundamental for trade, commerce and tourism. At the same time, they enable easy access to jobs, education, healthcare, and social activities, both in urban and rural regions. The status of road safety is crucial: harsh weather conditions can accelerate road degradation, and time the volume of traffic, including heavy traffic, grows, thus requiring frequent repairs. Any fault in maintenance could lead to severe incidents with death and injury worldwide. The latest report (2023) states that about 1.19 million people die each year as a result of road traffic crashes.

This paper explores a novel approach to traffic sign and road damage detection using advanced deep learning techniques, that are used in a Road Management System of a Public Municipality. This work is in progress within the Molise CTE research project, funded by the Italian Ministry of the Economic Growth (MIMIT), with the aim to leverage the best emerging technologies such as Cloud Computing, High Performance Computing, Artificial Intelligence, and AR/VR to develop and demonstrate state-of-the-art solutions in the Smart City environment.

The paper is structured as follows. Section 2 introduces a Literature Review, with description of more recent trends in research about the topic, then Section 3 provides a detailed description of the proposed solution, including information on the supporting cloud infrastructure. Section 4 gives a description of computational experiments with relevant metrics and giving evidence of the effectiveness of the proposed solution. Finally, Section 5 describes further improvements under development, the benefits and exploitation opportunities, and concludes the paper.

## 2. Literature Review

The detection and classification of traffic signs and road damage are vital components of intelligent transportation systems. Recent advancements in deep learning have significantly enhanced the capabilities of these systems, making them more accurate and efficient. This literature review explores various methods and approaches used in traffic sign detection, classification, and road damage detection, with a particular focus on the use of YOLO architecture and Convolutional Neural Networks (CNNs).

**Traffic Sign Detection and Classification**

1) ***Seminal Papers***: Early research laid the groundwork for the development of automated traffic sign detection and classification systems. An influential paper by Maldonado-Bascon et al. [2] introduces an automatic road-sign detection and recognition system utilizing Support Vector Machines (SVMs). The system improves driver-assistance by effectively detecting and recognizing various road sign shapes through a combination of color segmentation, linear SVM-based shape classification, and Gaussian kernel SVM recognition. The approach exhibits high success rates and robustness against transformations and occlusions. Another notable paper by Fang, Chen, and Fuh [3] outlines a method for road sign detection and tracking using neural networks and Kalman filters. This system leverages neural networks to extract color and shape features and employs Kalman filters to track the detected signs through image sequences, maintaining robust performance under diverse environmental conditions.
2) ***YOLO Architecture***: The YOLO (You Only Look Once) architecture has emerged as a popular choice for object detection tasks due to its real-time processing capabilities and high accuracy. Several studies have demonstrated the effectiveness of YOLO in traffic sign detection. For instance, Yang and Zhang (2020) [4] compared the performance of YOLOv4 and YOLOv3, finding that YOLOv4 significantly improved detection accuracy on a dataset of Chinese traffic signs. Similarly, Zhang (2023) [5] showed that YOLOv3 outperformed R-CNN algorithms in terms of speed and accuracy for traffic sign detection.

3) **Enhancements in YOLO**: Recent enhancements in YOLO include the development of lightweight models such as Sign-YOLO, which integrates the Coordinate Attention (CA) module and High-BiFPN to improve feature extraction and multi-scale semantic information fusion. This model achieved significant improvements in precision, recall, and detection speed on the CCTSDB2021 dataset [6]. Another example is the PVF-YOLO model, which uses Omni-Dimensional Convolution (ODconv) and Large Kernel Attention (LKA) to enhance detection accuracy and speed [7].

4) **Traffic Sign Classification**: After detection, traffic signs need to be classified into specific categories. Convolutional Neural Networks (CNNs) have been widely used for this purpose due to their strong feature extraction capabilities. Ciresan et al. (2012) [8] employed a CNN-based approach to classify German traffic signs, achieving state-of-the-art results. Additionally, improved YOLO models like TSR-YOLO have incorporated advanced modules to enhance accuracy in complex scenarios [9].

5) **Traffic Sign Damage Classification**: To the best of our knowledge, only two papers address traffic sign damage classification comprehensively. Ana Trpković, Milica Selmic, and Sreten Jevremović (2021) [10] developed a CNN model to identify and classify damaged and vandalized traffic signs. Another significant contribution is by J. N. Acilo et al. (2018) [11], who presented a study titled "Traffic Sign Integrity Analysis Using Deep Learning." This research employed transfer learning with the ResNet-50 architecture to classify the compliance and physical degradation status of traffic signs, achieving high accuracy.

**Road Damage Detection**

1) **Dataset Utilization**: Public datasets such as Mappilary **[18]** and the Road Damage Detection (RDD) **[19]** dataset provide comprehensive collections of annotated images for training and evaluating models. These datasets cover various types of road damage, including potholes, cracks, and surface wear, facilitating the development of robust detection models.

2) **Deep Learning Approaches**: Deep learning approaches, particularly those using CNNs and YOLO architectures, have shown significant advancements in road damage detection. Zhang et al. (2018) [12] employed a deep CNN model to detect road cracks from images, achieving high precision and recall rates. Similarly, Maeda et al. (2018) [13] applied YOLO to detect multiple types of road damage, demonstrating the model's effectiveness in real-world scenarios.

**GPS Data**

1) **Integration of GPS Data**: M. Strutu, G. Stamatescu, and D. Popescu (2013) [14] introduced a mobile sensor network-based system for monitoring road surfaces, incorporating 3D accelerometers, GPS, and video modules. Their research demonstrated the effectiveness of integrating multiple sensors for comprehensive road monitoring. Similarly, M. Perttunen et al. (2011) [15] developed a system for detecting road surface anomalies using accelerometers and GPS readings from mobile phones. Their pattern recognition system showcased the potential of mobile devices in monitoring road conditions effectively. Furthermore, R. Tarun and B. P. Esther (2023) [16] created an affordable road sign detection system utilizing a Raspberry Pi and GPS. Their system achieved high detection precision and demonstrated efficient real-time operation.

**Additional Insights from Recent Advances**

Recent advances in traffic sign recognition have explored a variety of machine learning and deep learning techniques. Lim et al. (2023) [17] provided a comprehensive overview of these advancements, highlighting the importance of preprocessing techniques, feature extraction methods, classification techniques, and the use of diverse datasets to address the challenges posed by different geographical regions, complex backgrounds, and varying illumination conditions.

Key contributions from recent studies include [17]:
- A comprehensive review of state-of-the-art traffic sign recognition work, categorizing studies into conventional machine learning and deep learning approaches.
- Discussion of widely adopted traffic sign recognition datasets, their challenges, and limitations, as well as future research prospects in this field.
- Emphasis on the importance of diverse datasets for improving model generalization and robustness.

## 3. Methodology

### First Datasets

#### Mapillary :

Mapillary Vistas Dataset **[18]** is a large-scale street-level imagery dataset designed for training and evaluating semantic segmentation models. This dataset is highly diverse, covering a wide range of environments, lighting conditions, and geographical locations. It includes various types of road signs, objects, and infrastructure commonly found in urban, suburban, and rural areas. The dataset is particularly useful for developing and testing algorithms for autonomous driving and urban planning applications.

- **Classes:** 401
- **Images:** 41,906
- **Size:** 32.8 GB

The dataset is split into training and validation sets, with 80% of the images used for training and 20% reserved for validation.

**RDD 2022:**

The Road Damage Detection (RDD) 2022 dataset **[19]** focuses on identifying and classifying different types of road surface damages. This dataset includes annotated images of road damage from various countries, making it a valuable resource for training machine learning models aimed at improving road maintenance and safety. The primary goal of using this dataset is to detect and classify road damages such as cracks, potholes, and other surface irregularities.

- **Classes:** 4
- **Images:** 34,007
- **Size:** 9.6 GB

Similar to the Mapillary dataset, the RDD 2022 dataset is also divided into an 80% training set and a 20% validation set. This split ensures that the models can be trained effectively while also being evaluated on a separate set of images to test their performance and generalization capabilities.

A comprehensive representation of the architecture for the road sign detection and classification system is depicted in Figure 1. For road damage, detection alone is sufficient, as these are already considered anomalies; therefore, the Yolo model is adequate.

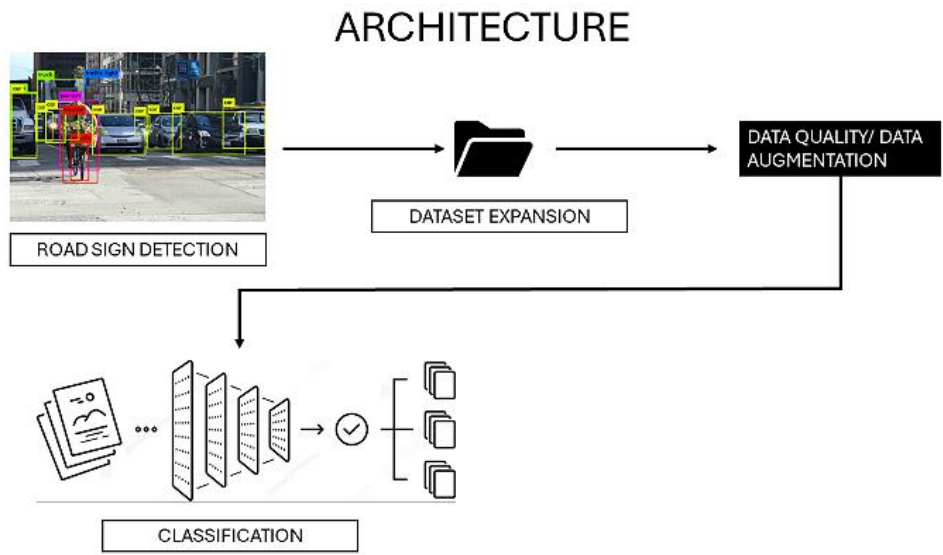

Fig. 1. Road Sign Workflow

*B. First Phase*

Before training the YOLO architectures, data manipulation is essential to ensure optimal performance and accuracy. The data manipulation techniques include the following:

- **Data Augmentation**: This involves applying various transformations to the training images, such as rotations, scaling, flipping, and color adjustments. These techniques help to increase the diversity of the training data and make the model more robust to different conditions.
- **Normalization:** Image pixel values are scaled to a standard range, typically between 0 and 1, to ensure uniformity and improve the convergence of the model during training.
- **Label Smoothing:** This technique is used to reduce overfitting by softening the hard labels in the training data, making the model less confident in its predictions and improving generalization.
- **Anchor Box Calculation:** Custom anchor boxes are computed based on the dataset to improve the detection accuracy of the YOLO model, especially for objects of various sizes.

## YOLOv8s for Road Surface Damages Detection:

YOLOv8s is a specific architecture within the YOLO (You Only Look Once) family, optimized for real-time object detection with a balance between speed and accuracy. It is particularly suitable for detecting road surface damages dueto its efficient design.

- **Pretrained: Yes**
- **Epochs:** 160
- **Image Size:** 640
- **Patience:** 100
- **Cache:** RAM
- **Device:** GPU
- **Batch Size:** 64

An example of a Road Damage detection is shown in Figure 2.

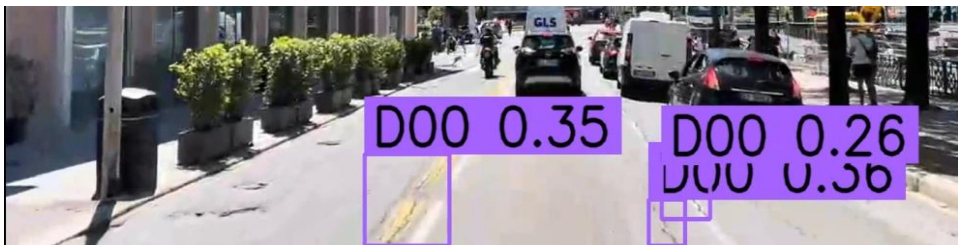

Fig. 2. Road damage detection

### YOLOv8x for Road Signs Detection:

YOLOv8x is a larger and more powerful version of the YOLO architecture, designed for detecting objects with higher precision and accuracy. This makes it well-suited for the detailed task of road signs detection.

- **Pretrained:** Yes
- **Epochs:** 100
- Image Size: 640
- **Patience:** 100
- **Cache:** RAM
- **Device:** GPU
- **Batch Size:** Auto

*C. Second Dataset*

Once the YOLO model is trained, it is necessary to build a dataset for the classification of road signs. To achieve this, videos recorded with a dashcam on the road are processed with YOLO, which crops the road signs from the frames. An example of the detected road sign is illustrated in Figure 3. The cropped signs are then labelled as damaged or not damaged based on the following criteria:
- Signs covered with spray-painted graffiti
- Signs covered with stickers
- Bent or physically damaged signs
- Rusty signs

The resulting dataset, in preliminary tests, is unbalanced, with **203 damaged** and **46 undamaged** signs. To address this imbalance, we employed two advanced techniques:

- **Focal Loss:** This loss function is designed to handle class imbalance by assigning more weight to hard-to-classify examples, reducing the impact of easily classified examples, and improving model performance on imbalanced data.
- **Cutout Regularization:** This technique involves randomly removing sections of the image during training. It helps improve model robustness and prevent overfitting, thereby enhancing the model's ability to generalize to new data.

These techniques allow us to effectively manage the dataset imbalance and improve the accuracy of classifying damaged and undamaged road signs .

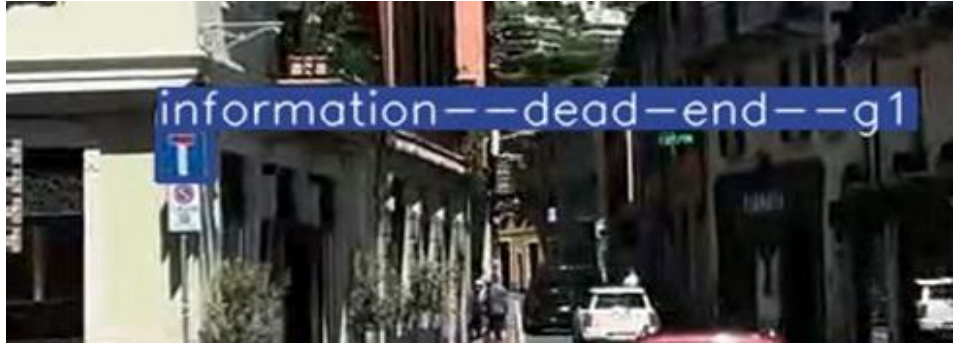

Fig. 3. Traffic sign detection

*D. Second Phase*

In this phase, we develop a Convolutional Neural Network (CNN) for classifying road signs as damaged or not damaged. The CNN architecture and training process are described as follows:

The CNN model consists of the following layers:

- An input layer that accepts images of size 150x150x3 (height, width, and color channels).
- A 2D convolutional layer with 32 filters of size 3x3, followed by a ReLU activation function.
- A max-pooling layer with a pool size of 2x2.
- A 2D convolutional layer with 64 filters of size 3x3, followed by a ReLU activation function.
- A max-pooling layer with a pool size of 2x2.
- A 2D convolutional layer with 128 filters of size 3x3, followed by a ReLU activation function.
- A max-pooling layer with a pool size of 2x2.
- A flattening layer to convert the 2D matrix data to a vector.
- A dense (fully connected) layer with 512 units and aReLU activation function.
- A dropout layer with a dropout rate of 0.5 to prevent overfitting.
- A dense output layer with 1 unit and a sigmoid activation function for binary classification.

The model is optimized by using Adam, with the Sigmoid Focal Cross-Entropy loss function, particularly effective for handling class imbalance. The accuracy metric is used to evaluate the model's performance.

To improve the robustness of the model and generalization capability, data augmentation techniques are applied, including rotation, width and height shifts, shear, zoom, and horizontal flips. Additionally, cutout regularization is implemented by randomly masking sections of the input images during training.

The training process involves:

- Training the model for 20 epochs with a batch size of 32.
- Using an 80-20 split for training and validation data.
- Evaluating the model's performance on validation set.
- Saving the trained model for future use.

The final evaluation on the validation set, in these first trials, shows a satisfactory level of accuracy **81%**, demonstrating the model's capability to classify road signs as damaged or not damaged effectively

## 4. Computational Experiments

*A. Computational Characteristics*

The training of the YOLO models is conducted using Google Colab, leveraging the NVIDIA Tesla T4 GPU. Google Colab provides a high-performance computing environment suitable for deep learning tasks. The key specifications of the hardware used are as follows:

- **GPU:** NVIDIA Tesla T4
    - CUDA Cores: 2560
    - Tensor Cores: 320
    - GPU Memory: 16 GB GDDR6
    - Memory Bandwidth: 320 GB/s
    - Performance: Up to 8.1 TFLOPS (FP32)
- **CPU:** Intel(R) Xeon(R) CPU
    - vCPUs: 2 (Base Frequency: 2.3 Ghz)
- **RAM:** 12.7 GB available in the Colab environment
- **Disk:** 100 GB available storage

For the CNN, training was performed on Reevo servers from Tiscali with the following specifications:

- **RAM:** 32 GB
- **CPU:** 24 vCPUs (Base Frequency: 2.5 Ghz)

The combination of these computational resources provides a robust environment for training and validating the deep learning models, enabling efficient processing of large datasets and complex computations required for road sign detection and classification.

*B. Metrics for Performance Evaluation and Results*

To evaluate the performance of the YOLOv8x model for road sign detection, we analyze several key metrics, including accuracy, box loss, and object loss. These metrics provide insights into the model's effectiveness in detecting and classifying road signs accurately.

*1) YOLOv8x Accuracy:* Figure 4 shows the accuracy of the YOLOv8x model over the training epochs. The accuracy metric includes the mean Average Precision (mAP) at different Intersection over Union (IoU) thresholds and other performance metrics such as precision and recall.

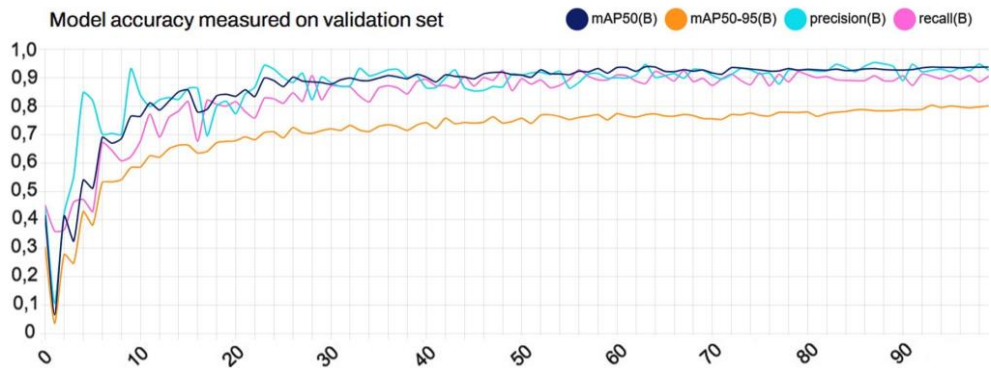

Fig. 4. YOLOv8x Accuracy

Specifically, we track:

- **mAP50:** Mean Average Precision at 50% IoU threshold.
- **mAP50-95:** Mean Average Precision averaged over IoU thresholds from 50
- **Precision:** The ratio of true positive detections to the total number of positive detections (true positives + false positives).
- **Recall:** The ratio of true positive detections to the total number of actual positives (true positives + false negatives).

The graph indicates the following trends:

- The **mAP50** metric (blue line) shows a steady improvement, stabilizing around 0.9, indicating a high level of accuracy for the model in detecting objects with a 50% IoU threshold.
- The **mAP50-95** metric (orange line) improves gradually, reflecting the model's performance across a wider range of IoU thresholds. It stabilizes around 0.7, showcasingthe model's robustness in varying detection conditions.
- **Precision** (cyan line) shows fluctuations but generally trends upwards, indicating improvements in the model's ability to reduce false positives over time.
- **Recall** (magenta line) also improves and stabilizes around 0.8, demonstrating the model's effectiveness in capturing most of the actual positive instances.

*2) YOLOv8x Box Loss:* Figure 5 illustrates the box loss during training. Box loss measures the error in predicting the bounding boxes for detected objects. It is a crucial metric for object detection models as it directly affects the precision of the detected objects' locations. Lower box loss values indicate more accurate predictions of the bounding box coordinates.

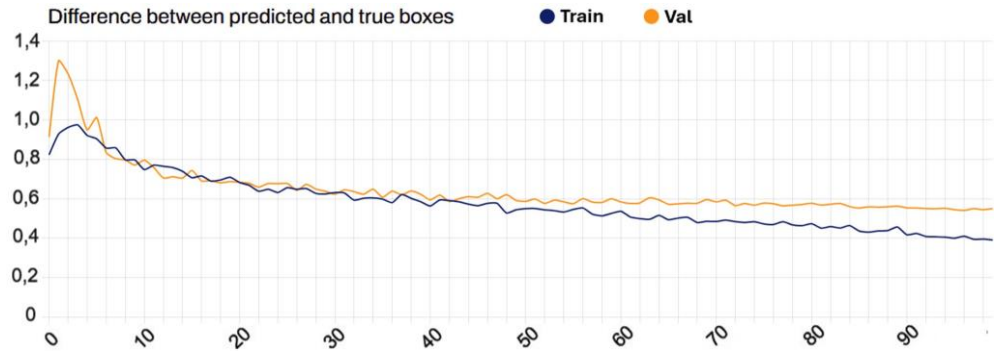

Fig. 5. YOLOv8x Box Loss

The graph shows that the box loss decreases significantly during the initial epochs and stabilizes over time, indicating that the model is learning to accurately predict the bounding box locations of the road signs.

*3) YOLOv8x Object Loss:* Figure 6 presents the object loss over the training epochs. Object loss evaluates the error in classifying whether a particular region in the image contains an object of interest. Lower object loss values reflect the model's enhanced capability to distinguish between objects and the background, leading to more reliable detections.

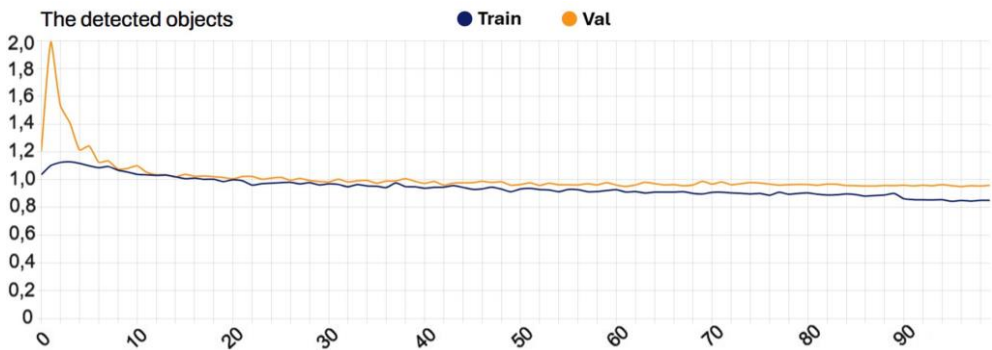

Fig. 6. YOLOv8x Object Loss

The graph indicates that the object loss decreases rapidly during the initial epochs and then gradually stabilizes, suggesting that the model becomes increasingly proficient at distinguishing between road signs and background noise as training progresses.

These metrics collectively provide a comprehensive overview of the YOLOv8x model's performance in detecting and classifying road signs. The continuous improvement in accuracy and reduction in both box and object loss throughout the training process indicate the model's effectiveness and robustness in handling the task of road sign detection.

A similar analysis applies to YOLOv8s, which is used for detecting road surface damages. The following observations were made:

**Accuracy:** The accuracy metrics for YOLOv8s, as shown in Figure 7, indicate steady improvement over epochs, with metrics such as mAP50, mAP50-95, precision, and recall showing consistent performance gains.

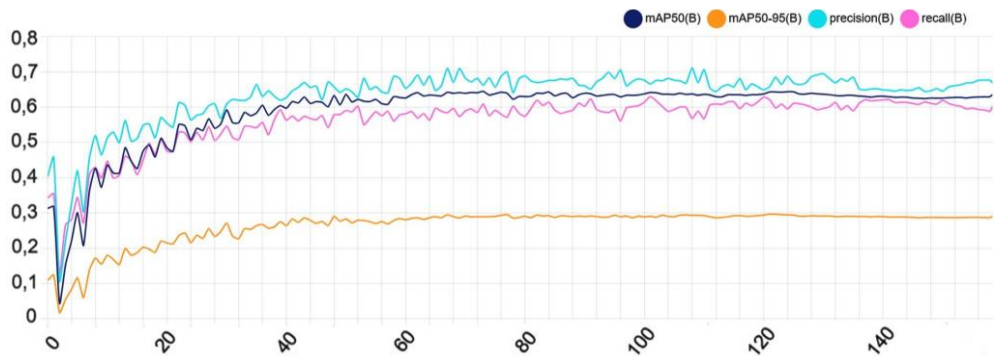

Fig. 7. YOLOv8s Accuracy

- **Box Loss:** As shown in Figure 8, the box loss decreases over time, indicating improved precision in predicting thebounding box locations for road damages.

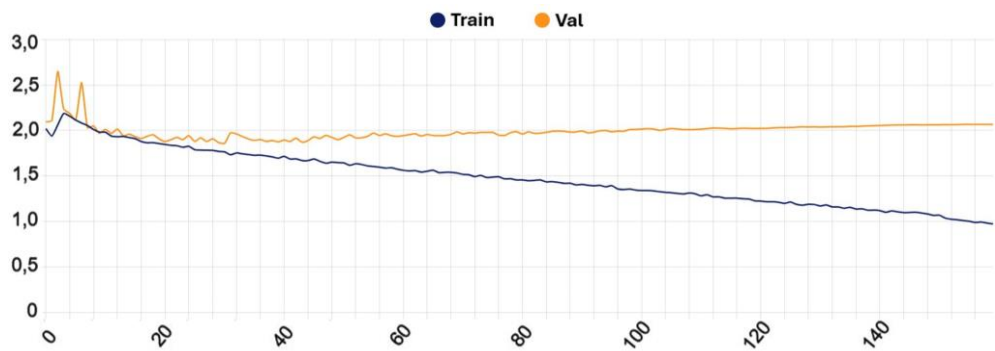

Fig. 8. YOLOv8s Box Loss

- **Object Loss:** The object loss, depicted in Figure 9, shows a downward trend, demonstrating enhanced capability in distinguishing between damaged and undamaged road surfaces.

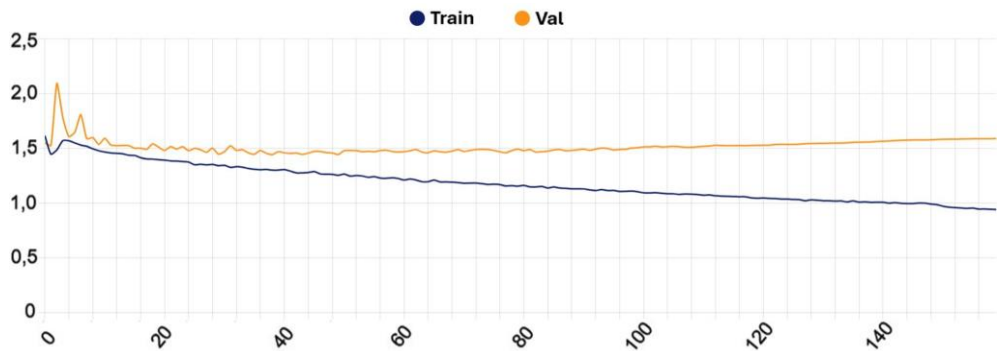

Fig. 9. YOLOv8s Object Loss

These metrics collectively illustrate that YOLOv8s is effective in detecting and classifying road surface damages, showing consistent improvement in accuracy and a reduction in both box and object loss across the training epochs.

*4) CNN:* Graph 10 indicates the following trends:

- **High Training Accuracy and Precision:** The training accuracy metric (blue line) starts low and increases rapidly within the first few epochs. After the initial increase, the accuracy stabilizes around a high value, indicating that the model is effectively learning the training data and making correct predictions and positive identifications. Training precision (green line) remains relatively high, though it fluctuates, suggesting some variability in the learning process regarding positive predictions.
- **Validation Metrics**: The validation accuracy metric (orange line) starts low and increases over the first few epochs, stabilizing at a value slightly lower than the training accuracy. This indicates that the model is generalizing reasonably well to the validation set without overfitting too much. Validation precision (red line) shows significant fluctuation throughout the epochs, likely due to dataset imbalance and its limited size, which is logical in this preliminary testing phase. Both training and validation recall (purple and pink lines, respectively) are high and fairly stable, indicating that the model is effectively identifying positive instances with minimal false negatives.
- **Potential Overfitting**: There is a slight indication of overfitting due to the gap between training and validation precision. This overfitting is likely attributable to the small size and imbalance of this preliminary phase, as other influencing factors such as model complexity, data augmentation, and hyperparameter tuning (e.g., regularization, epochs) have been appropriately addressed.

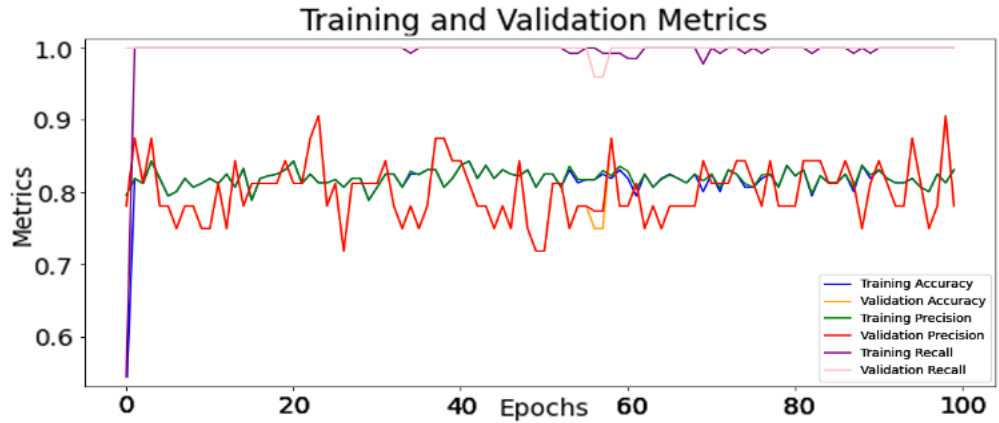

Fig. 10. CNN metrics

## 5. Conclusions

In this study, we successfully develop and train YOLO models for road sign detection and CNN models for classifying road signs as damaged or not damaged. Our approach utilizes data augmentation and cutout regularization techniques to improve the robustness and generalization of our models. The computational experiments conducted on Google Colab and Tiscali servers demonstrates the effectiveness of our methods in handling large datasets and complex computations.

For future work, we propose the following extensions to enhance the capabilities and applications of our models:

- **Incorporating Retroreflectivity Factors:** To further refine the classification of road signs, we plan to include retroreflectivity factors in our analysis. This involves detecting and classifying faded or discolored signs, which can significantly impact road safety. Developing models that can identify such signs will be crucial for timely maintenance and replacement.
- **Leveraging Generative AI for Data Labeling:** The process of manually labeling large datasets is time- consuming and prone to human error. By employing generative AI techniques, we can automate the labeling process, thereby reducing the time and effort required. This will also enable us to handle larger datasets more efficiently.
- **Generating Synthetic Data for Balanced Datasets:** One of the challenges we faced was the imbalance between damaged and non-damaged road signs in our dataset. To address this, we propose using generative AI to create synthetic images of damaged signs. By artificially "damaging" images of non-damaged signs (e.g., adding graffiti, stickers, rust, and physical damage), we can construct a balanced dataset. This synthetic data will enhance the training process, making our models more robust and accurate.

By implementing these extensions, we aim to improve the accuracy and reliability of road sign detection and classification systems. This will contribute to better road safety and maintenance practices, ultimately benefiting road users and maintenance authorities.

**Acknowledgments**   This work is supported by the Molise CTE Project, funded by MIMIT (FSC 2014- 2020), grant #D33B22000060001.

We would like to extend our gratitude to the project coordinators and funding bodies for their support and resources, which were instrumental in the successful completion of this research.